\documentclass[12pt,preprint]{aastex}

\newcommand{\be}{\begin{equation}}
\newcommand{\ee}{\end{equation}}
\newcommand{\ba}{\begin{eqnarray}}
\newcommand{\ea}{\end{eqnarray}}

\shorttitle{Reduced convergence and the local smoothness parameter}
\shortauthors{Wang et al.}

\begin{document}

\title{Reduced convergence and the local smoothness parameter:
bridging two different descriptions of weak lensing amplification}
\author{Yun~Wang$^{1}$, Jason Tenbarge$^{1}$, and Bobby Fleshman$^{1}$\\
                 {\it (Draft version, December 31, 2004)} }
\altaffiltext{1}{Department of Physics \& Astronomy, Univ. of Oklahoma,
                 440 W Brooks St., Norman, OK 73019;
                 email: wang@nhn.ou.edu}

\begin{abstract}

Weak gravitational lensing due to the inhomogeneous matter distribution 
in the universe is an important systematic uncertainty in the use of 
standard candles in cosmology.
There are two different descriptions of weak lensing amplification,
one uses a local smoothness parameter $\tilde{\alpha}$, the
other uses reduced convergence $\eta= 1+ \kappa/|\kappa_{min}|$
(where $\kappa$ is convergence).
The $\tilde{\alpha}$ description involves 
Dyer-Roeder distance $D_A(\tilde{\alpha}|z)$ ($\tilde{\alpha}=1$
corresponds to a smooth universe); it is simple and convenient,
and has been used by the community to illustrate the
effect of weak lensing on point sources such as
type Ia supernovae. 
Wang (1999) has shown that the $\tilde{\alpha}$ description
can be made realistic by allowing $\tilde{\alpha}$ to be
a local variable, the local smoothness parameter.
The $\eta$ description has been used by
Wang, Holz, \& Munshi (2002) to derive a universal probability 
distribution (UPDF) for weak lensing amplification. 
In this paper, we bridge the two different descriptions
of weak lensing amplification by relating 
the reduced convergence $\eta$ and the local smoothness parameter
$\tilde{\alpha}$. We give the variance of $\tilde{\alpha}$
in terms of the matter power spectrum, thus providing
a quantitative guidance to the use of Dyer-Roeder distances
in illustrating the effect of weak lensing.
The by-products of this work include a corrected
definition of the reduced convergence, and
simple and accurate analytical expressions
for $D_A(\tilde{\alpha}|z)$.
Our results should be very useful 
in studying the weak lensing of standard candles.

\end{abstract}



\section{\label{sec:level1}Introduction}

Type Ia supernovae (SNe Ia)
are our best candidates for cosmological standard candles
\citep{Phillips,Riess95}, and have lead to the startling
discovery that the expansion of our universe is accelerating
\citep{Schmidt98,Riess98,Perl99}.
This observational fact can best be explained by
the existence of dark energy in the universe \citep{Garna98,SNfluxavg}.
It is important that further and more ambitious SN Ia observations
are carried out \citep{SNpencilbeam,Wang01}.
There are a number of ambitious SN Ia programs that are
active or planned. 
The CFH Legacy Survey, the ESSENCE project at NOAO, and the 
proposed Large Synoptic Survey Telescope (LSST) (http://www.lsst.org/)
and Supernova / Acceleration Probe (SNAP) (http://snap.lbl.gov), will
give us an impressive number of supernovae.
The data from such
observations must be properly analyzed with good understanding of
the systematic uncertainties.

Weak gravitational lensing due to the inhomogeneous matter distribution 
in the universe is an important systematic uncertainty in the use of 
standard candles in cosmology 
\citep{Kantow95,frieman97,Wamb97,HolzWald98,Wang99,Valageas00,MJ00,Barber00,Premadi01}.
Both the smoothness parameter $\tilde{\alpha}$
\citep{Kantow95,Wang99,Sereno01,Sereno02}
and the reduced convergence $\eta$ \citep{Valageas00,MJ00,MunshiWang03}
have been used in the literature
to study the weak lensing of standard candles.

Magnification of a standard candle 
can be expressed in terms of the ratio of our actual distance to the
standard candle and our distance to it in a completely smooth
universe. Such distances, $D_A(\tilde{\alpha}|z)$, are solutions 
to the Dyer-Roeder equation \citep{DR73,Kantow98},
with the matter inhomogeneity parametrized by a local smoothness
parameter $\tilde{\alpha}$ \citep{Wang99}.
Due to their simplicity and convenience, Dyer-Roeder distances
with fixed values of $\tilde{\alpha}$ are
used to illustrate the effect of weak lensing of point sources,
although they do not correctly describe weak lensing \citep{Hamana}.
However, since the Dyer-Roeder equation follows from the general 
equation for light deflection in general relativity 
(for a given mass density) \citep{Sch92},
it is valid when shear can be neglected (as is the case for weak lensing
amplification), as long as mass density is not fixed as in the usual
applications of the Dyer-Roeder equation. 
In \cite{Wang99}, a new and valid way of using the Dyer-Roeder 
equation has been developed: by allowing
a dispersion in the possible values of
$\tilde{\alpha}$ (with probability density function derived
from the matter power spectrum) due to the inhomogeneous distribution of matter,
we can very efficiently compute the lensing effect and 
intuitively understand the weak lensing of cosmological standard
candles.

The reduced convergence $\eta$ has been used by
Wang, Holz, \& Munshi (2002) to derive a universal probability 
distribution (UPDF) for weak lensing amplification. 
To achieve a unified understanding of weak lensing magnification,
it is important that we understand 
how the local smoothness parameter $\tilde{\alpha}$ and
the reduced convergence $\eta$ are related.

This paper is organized as follows.
We derive analytical solutions to the Dyer-Roeder equation 
in Sec.II, to be used in the rest of the paper. 
In Sec.III, we relate the local smoothness parameter $\tilde{\alpha}$
and the reduced convergence $\eta$.
In Sec.IV, we derive the variance of $\tilde{\alpha}$ in terms 
of matter power spectrum.
Sec.V contains a summary and discussions.
The Appendix contains simple fitting formulae that comprise
an accurate expression of the Dyer-Roeder distance.

\section{Angular diameter distance as function of the smoothness 
parameter $\tilde{\alpha}$}
\label{app:DAalpha}

In a Hubble diagram of standard candles, one must use
distance-redshift relations to make theoretical interpretations. 
The distance-redshift relations depend on the distribution of
matter in the universe. 
\cite{Wang99} has shown that we can define the local or
{\it direction dependent} smoothness parameter $\tilde{\alpha}$
via the Dyer-Roeder equation \citep{DR73}; there is a unique mapping between 
$\tilde{\alpha}$ and the magnification of a source.
In this section, we express the angular diameter
distance to a standard candle in terms of $\tilde{\alpha}$.
This will be used in the next section to relate $\tilde{\alpha}$
to the reduced convergence $\eta$.

The local smoothness parameter $\tilde{\alpha}$ essentially represents
the amount of matter that can cause magnification of a given source.
Since matter distribution in our universe is inhomogeneous, we can think
of our universe as a mosaic of cones centered on the observer, 
each with a different value of $\tilde{\alpha}$.

In a smooth Friedmann-Robertson-Walker (FRW) universe,
$\tilde{\alpha}=1$ in all beams;
the metric is given by $ds^2=dt^2-a^2(t)[dr^2/(1-kr^2)+r^2 (d\theta^2
+\sin^2\theta \,d\phi^2)]$, where $a(t)$ is the cosmic scale factor,
and $k$ is the global curvature parameter ($\Omega_k
=1-\Omega_m-\Omega_X=-k/H_0^2$).
The comoving distance $r$ is given by \citep{Weinberg72}
\ba
\label{eq:r(z)}
r(z) &=&\frac{cH_0^{-1}}{|\Omega_k|^{1/2}}\,
\mbox{sinn}\left\{ |\Omega_k|^{1/2}
\int_0^z dz'\,\frac{1}{E(z')} \right\}, \nonumber\\
E(z) &\equiv& \sqrt{\Omega_m (1+z)^3 +\Omega_X f_X(z) + \Omega_k (1+z)^2},
\ea
where ``sinn'' is defined as sinh if $\Omega_k>0$, and sin if  $\Omega_k<0$.
If $\Omega_k=0$, the sinn and $\Omega_k$'s disappear from Eq.(\ref{eq:r(z)}),
leaving only the integral.
$\Omega_X f_X(z)$ is the contribution from dark energy;
the dimensionless dark energy density $f(z)=\rho_X(z)/\rho_X(z=0)$.
For a cosmological constant, $\Omega_X=\Omega_\Lambda$, and $f_X(z)=1$. 
The angular diameter distance is given by $d_A(z)=r(z)/(1+z)$,
and the luminosity distance is given by $d_L(z)=(1+z)^2 d_A(z)$.

However, our universe is clumpy rather than smooth.
According to the focusing theorem in gravitational lens theory,
if there is any shear or matter along a beam connecting 
a source to an observer, the angular diameter distance of
the source from the observer is {\it smaller} than that which would occur
if the source were seen through an empty, shear-free cone,
provided the affine parameter distance (defined such that its element
equals the proper distance element at the observer) 
is the same and the beam has not gone through a caustic.
An increase of shear or matter density along the beam decreases the
angular diameter distance and consequently increases the
observable flux for given $z$. \citep{Sch92} 
For studies of weak lensing magnification (with convergence
$|\kappa| \la 0.2$ \citep{Barber00}), we can ignore
shear and consider convergence only, which corresponds
to the matter in the beam.

For given redshift $z$,
if a mass-fraction $\tilde{\alpha}$ of the matter in the universe is smoothly
distributed, the largest possible distance for
light bundles which have not passed through a caustic is given by
the Dyer-Roeder distance,
the solution to the Dyer-Roeder equation \citep{DR73,Kantow98}:
\ba
\label{eq:DR}
&& g(z) \, \frac{d\,}{dz}\left[g(z) \frac{dD_A}{dz}\right]
+\frac{3}{2} \tilde{\alpha} \,\Omega_m (1+z)^5 D_A=0, \nonumber \\
&& D_A(z=0)=0, \hskip 1cm \left.\frac{dD_A}{dz}\right|_{z=0}=\frac{c}{H_0},
\nonumber \\
&& g(z) \equiv  (1+z)^2\, E(z).
\ea


The angular diameter distance for a given smoothness parameter $\tilde{\alpha}$
and redshift $z$, $D_A(\tilde{\alpha}|z)$, can be obtained via the 
numerical integration of Eq.(\ref{eq:DR}). However, it is instructive to express
$D_A(\tilde{\alpha}|z)$ in terms of $\tilde{\alpha}$.

Integrating Eq.(\ref{eq:DR}) gives
\be
g(z) \, \frac{d D_A}{dz} - cH_0^{-1} = - \frac{3}{2} \Omega_m \tilde{\alpha}
\int_0^z \frac{(1+z')^5}{g(z')}\, D_A(\tilde{\alpha}|z')\ dz',
\ee
where we have used the boundary condition on $d D_A/dz$ at $z=0$.
Integrating the above equation gives
\be
D_A(\tilde{\alpha}|z) = cH_0^{-1}\,\int_0^z \frac{dz'}{g(z')}
- \frac{3}{2} \Omega_m \tilde{\alpha}
\int_0^z \frac{dz'}{g(z')}\, \int_0^{z'}
\frac{(1+z'')^5}{g(z'')}\, D_A(\tilde{\alpha}|z'')\ dz''.
\ee
Now we exchange the order of integration in the double integral above,
i.e., $\int_0^z dz' \int_0^{z'} dz'' = \int_0^z d z'' \int_{z''}^z dz'$.
Hence
\be
D_A(\tilde{\alpha}|z) = cH_0^{-1}\,\int_0^z \frac{dz'}{g(z')}
- \frac{3}{2} \Omega_m \tilde{\alpha}
\int_0^z \frac{(1+z'')^5}{g(z'')}\, D_A(\tilde{\alpha}|z'')\ dz'' \,
\int_{z''}^{z}\frac{dz'}{g(z')}.
\ee

Now we have
\be
\label{eq:DR2}
D_A(\tilde{\alpha}|z) = - \frac{3}{2} \, \Omega_m \,\tilde{\alpha}
\int_0^z dz'\, \frac{(1+z')^5}{g(z')}\, D_A(\tilde{\alpha}|z')
\,\frac{ [\lambda(z)-\lambda(z')]}{cH_0^{-1}}
+ D_A(\tilde{\alpha}=0|z),
\ee
where the affine parameter 
\be
\lambda(z) = cH_0^{-1}\, \int_0^z \frac{dz'}{g(z')}.
\ee
Note that $D_A(\tilde{\alpha}=0|z)=\lambda(z)$.
\cite{SNfluxavg} has shown that $D_A(\tilde{\alpha}=1|z)=r(z)/(1+z)$,
where $r(z)$ is the comoving distance in a smooth universe.

No approximations have been made in obtaining Eq.(\ref{eq:DR2})
from the Dyer-Roeder equation Eq.(\ref{eq:DR}).
We can solve the Eq.(\ref{eq:DR2}) perturbatively by replacing
$D_A(\tilde{\alpha}|z')$ in the integral on the right hand side 
with its previous order approximation.
In the first order perturbation, we replace $D_A(\tilde{\alpha}|z')$
in the integral by its 0th order approximation,
$D_A^{(0)}(\tilde{\alpha}|z')=D_A(\tilde{\alpha}=1|z')$.
This gives us
\be
D_A^{(1)}(\tilde{\alpha}|z)= D_A(\tilde{\alpha}=1|z)\, \tilde{\alpha} 
\, \tilde{\kappa}_{min}(z) + D_A(\tilde{\alpha}=0|z)
=D_A(\tilde{\alpha}=1|z) \, \left[ 1+ \tilde{\kappa}_{min}(z) 
(\tilde{\alpha} -1 ) \right],
\ee
where we have defined
\be
\label{eq:kapdef}
\tilde{\kappa}_{min}(z) \equiv 1 - \frac{D_A(\tilde{\alpha}=0|z)}
{D_A(\tilde{\alpha}=1|z)}
= - \frac{3}{2} \, \frac{\Omega_m}{cH_0^{-1}}
\int_0^z dz'\, \frac{(1+z')^2}{E(z')}\, \frac{r(z')}{r(z)}\, (1+z)
\,[\lambda(z)-\lambda(z')],
\ee
using $D_A(\tilde{\alpha}=1|z)=r(z)/(1+z)$ \citep{SNfluxavg}.
Note that $\tilde{\kappa}_{min}(z) <0 $.

After four more iterations, we find the fifth order perturbative solution
\be
\label{eq:exp}
\frac{D_A(\tilde{\alpha}|z)}{D_A(\tilde{\alpha}=1|z) }
= 1 - |\tilde{\kappa}_{min}(z)| (\tilde{\alpha}-1)\,
\left\{ 1- \tilde{\alpha} \left[ C_1(z) - C_2(z)\, \tilde{\alpha}
+ C_3(z) \tilde{\alpha}^2 - C_4(z) \tilde{\alpha}^3
+ {\cal O}(\tilde{\alpha}^4) \right] \right\},
\ee
The functions $C_i(z)$ ($i=1,2,3,4$) are given by
\be
\label{eq:A,B,C,D}
C_i(z) \equiv \frac{3}{2} \, \frac{\Omega_m}{|\tilde{\kappa}_{min}(z)|}
\int_0^z dz'\, \frac{(1+z')^3}{E(z')}\, \frac{D_A(\tilde{\alpha}=1|z')}
{D_A(\tilde{\alpha}=1|z)}\,\left|\tilde{\kappa}_{min}(z')\right|\, C_{i-1}(z')\,
\frac{ [\lambda(z)-\lambda(z')]}{cH_0^{-1}},
\ee
where we have defined $C_0 \equiv 1$.

Fig.1 shows the difference between the angular diameter distance given
by the solution of the Dyer-Roeder equation and our analytical expansion
to 4th order in $\tilde{\alpha}$ (see Eq.(\ref{eq:exp})), with
$\tilde{\kappa}_{min}(z)$, and $C_i(z)$ ($i=1,2,3,4$)
computed numerically using Eq.(\ref{eq:A,B,C,D}).
Clearly, the accuracy of Eq.(\ref{eq:exp}) is practically unlimited,
as additional terms can be added in a straightforward manner as needed.

Note that we have not yet made {\it any} assumptions about dark energy;
the results of this subsection are valid for {\it all}
cosmological models. 
Simple and accurate fitting formulae for 
$\tilde{\kappa}_{min}(z)$, and $C_i(z)$ ($i=1,2,3,4$) are given in the Appendix.

\section{Reduced convergence and the local smoothness
parameter $\tilde{\alpha}$}

\cite{Wang02} have derived a universal distribution  
for the weak lensing magnification distribution of standard candles, 
valid for all cosmological models, with arbitrary matter
distributions, over all redshifts. Their results are based on
a universal probability distribution function (UPDF),
$P(\eta)$, for the reduced convergence, $\eta$.  For a given
cosmological model, the magnification probability
distribution, $P(\mu)$, at redshift $z$ is related to the
UPDF by $P(\mu)=P(\eta)/2\left|\kappa_{min}\right|$, where
$\eta=1+(\mu-1)/(2|\kappa_{min}|)$, and
$\kappa_{min}$ (the minimum convergence) can be directly
computed from the cosmological parameters.
The UPDF can be well
approximated by a three-parameter stretched Gaussian
distribution, where the values of the three parameters
depend only on $\xi_\eta$, the variance of $\eta$. In short,
all possible weak lensing probability distributions can be
well approximated by a one-parameter family. 
Each alternative cosmological model is then
described by a single function $\xi_\eta(z)$.  
This method gives $P(\mu)$ in excellent agreement with
numerical ray-tracing and three-dimensional shear matrix 
calculations, and provide numerical
fits for three representative models (SCDM, $\Lambda$CDM,
and OCDM).  

It is of interest to relate the local smoothness
parameter $\tilde{\alpha}$ to the reduced convergence
$\eta$, as both can, and have been used to
parametrize weak lensing of standard candles.

The magnification factor of a source is given precisely by \citep{Sch92}
\be 
\label{eq:mugam}
\mu = \frac{1}{(1-\kappa)^2 -\gamma^2},
\ee
where $\kappa$ is convergence (which depends only on matter within
the beam), and $\gamma$ is shear (which depends on matter outside
of the beam). $\kappa$ and $\gamma$ can be computed from the
deflection potential.
Ignoring shear, we can relate magnification $\mu$ to convergence $\kappa$:
\be 
\mu = \frac{1}{(1-\kappa)^2},
\label{eq:mukap}
\ee
which is valid for $|\kappa| \la 0.2$ (equivalent to $\mu \la 1.56$)
for all cosmological models at all redshifts \citep{Barber00}.

In terms of angular diameter distances, 
magnification can be defined as \citep{Sch92}
\be 
\label{eq:muDAalpha}
\mu \equiv \left| \frac{D_A(\tilde{\alpha}=1)}{D_A(\tilde{\alpha})}
\right|^2.
\label{eq:muDA}
\ee

Comparing Eqs.(\ref{eq:mukap}) and (\ref{eq:muDA}), we have 
\be
\kappa = 1- \frac{D_A(\tilde{\alpha})}{D_A(\tilde{\alpha}=1)}.
\ee
The {\it minimum convergence} corresponds to the 
{\it minimum magnification} $\mu_{min}$, which
occurs when $\tilde{\alpha}=0$
(empty beam, when the smoothly distributed matter fraction in the beam is zero),
i.e., $\rho=0$, or $\delta=(\rho- \overline{\rho})/\overline{\rho} = -1$.
Since the minimum magnification can only be achieved when $\gamma=0$
(see Eq.(\ref{eq:mugam})), the true minimum convergence 
is given precisely by 
\be
\tilde{\kappa}_{min} = 1- \frac{D_A(\tilde{\alpha}=0)}{D_A(\tilde{\alpha}=1)},
\ee
as we have defined in the previous section.
Note that the smallest possible magnification, $\mu_{min}$, 
corresponds to $\tilde{\kappa}_{min}$ by definition,
\be
\label{eq:mumin}
\mu_{min} = \left| \frac{D_A(\tilde{\alpha}=1)}{D_A(\tilde{\alpha}=0)}\right|^2
= \frac{1}{(1-\tilde{\kappa}_{min})^2}.
\ee

In the context of weak lensing, $\kappa$ is the projected density contrast,
and $\kappa_{min}$ is the projected density contrast when $\delta=-1$
along the line of sight from observer to source \citep{Valageas00,MJ00}:
\ba
\kappa_{min} &\equiv & - \frac{3}{2} \, \frac{\Omega_m}{(cH_0^{-1})^2}
\int_0^{\chi} d{\chi}'\, \frac{r(\chi') r(\chi-\chi')}{r(\chi)}\, (1+z')\,
\nonumber \\
d\chi & = & \frac{cH_0^{-1}\, dz}{E(z)},
\hskip 1cm
r(\chi)= \frac{CH_0^{-1}}{\sqrt{ |\Omega_k|}} 
\mbox{sinn}\left( \sqrt{ |\Omega_k|}\, \chi \right).
\label{eq:dchi}
\ea
Thus
\be
\frac{\tilde{\kappa}_{min}(z)}{\kappa_{min}(z)}
=\frac{ \int_0^z dz'\, \frac{(1+z')^2}{E(z')}\, r(z')\, (1+z)
\,[\lambda(z)-\lambda(z')]}
{\int_0^z dz'\, \frac{(1+z')}{E(z')}\, r(z')\, r(\chi-\chi')},
\ee
where we have used Eq.(\ref{eq:kapdef}).
Fig.2 shows the difference between $\tilde{\kappa}_{min}(z)$
and $\kappa_{min}(z)$ for three representative cosmological models.

To see why $\kappa_{min}$ differs from $\tilde{\kappa}_{min}$, let us write
them in more intuitive forms:
\be
\tilde{\kappa}_{min}(z) = 
- \frac{3}{2} \, \frac{\Omega_m}{cH_0^{-1}}
\int_0^z dz'\, \frac{(1+z')^2}{E(z')}\, \frac{D_A(\tilde{\alpha}=1|z')}
{D_A(\tilde{\alpha}=1|z)}\, D_A(\tilde{\alpha}=0|z',z),
\ee
where we have used $D_A(\tilde{\alpha}=1|z)=r(z)/(1+z)$ \citep{SNfluxavg},
and $D_A(\tilde{\alpha}=0|z',z)= (1+z')\, [\lambda(z)-\lambda(z')]$
is the angular diameter distance of a source at $z$ from a fictitious
observer at $z'$ for $\tilde{\alpha}=0$ \citep{Sch92}.
Similarly, we can write
\be
\kappa_{min}(z) = 
- \frac{3}{2} \, \frac{\Omega_m}{cH_0^{-1}}
\int_0^z dz'\, \frac{(1+z')^2}{E(z')}\, \frac{D_A(\tilde{\alpha}=1|z')}
{D_A(\tilde{\alpha}=1|z)}\, \frac{r \left( \chi(z)-\chi(z') \right)}{1+z},
\ee
where $r\left( \chi(z)-\chi(z')\right)/(1+z)$ is the angular diameter
distance between $z$ and $z'$ in a completely smooth universe 
(i.e., filled beam, $\tilde{\alpha}=1$).
Since the true minimum convergence should correspond to the minimum magnification,
which occurs when the beam connecting the source and observer is
{\it empty} (i.e., $\tilde{\alpha}=0$), $\tilde{\kappa}_{min}(z)$
is the true minimum convergence, while $\kappa_{min}(z)$ is 
its approximation when the filled beam (instead of the
empty beam) angular diameter distance is used along the line of sight
from the observer to the source.

$\kappa_{min}(z)$ has been mistakenly associated with $\mu_{min}$
in the literature. Note that $|\kappa_{min}| <  |\tilde{\kappa}_{min}|$,
hence $1/(1-\kappa_{min})^2$ underestimates the minimum magnification
$\mu_{min}$.
Some of the consequences of the subtle
difference between $\tilde{\kappa}_{min}$ and $\kappa_{min}$
will be presented in Sec.IV.

Fig.2 shows that the correction to $\kappa_{min}(z)$ required for
the accurate calculation of the minimum convergence 
$\tilde{\kappa}_{min}(z)$ is quite modest (less than 2\% at $z\leq 1$)
in cosmological models that fit current observational data (the 
$\Lambda$CDM model).

Let us define a modified reduced convergence given by
\be
\label{eq:etatil}
\tilde{\eta} \equiv 1 + \frac{\kappa}{|\tilde{\kappa}_{min}|},
\ee
which should be compared with the reduced convergence defined
by \cite{Valageas00} and \cite{MJ00},
\be
\eta \equiv 1 + \frac{\kappa}{|{\kappa}_{min}|}
\ee
$\eta$ is related to $\tilde{\eta}$ by
\be
\eta=1 + \left[\tilde{\eta} -1 \right] 
\frac{|\tilde{\kappa}_{min}|}{|{\kappa}_{min}|}.,
\ee

Since
\be
\kappa- \tilde{\kappa}_{min} = \frac{D_A(\tilde{\alpha}=0)-
D_A(\tilde{\alpha})}{D_A(\tilde{\alpha}=1)},
\ee
we find that
\be
\label{eq:etatil-alpha}
\tilde{\eta} = \tilde{\alpha}\,
\left\{ 1- (\tilde{\alpha}-1) \left[ C_1(z) - C_2(z)\, \tilde{\alpha}
+ C_3(z) \tilde{\alpha}^2 - C_4(z) \tilde{\alpha}^3
+ {\cal O}(\tilde{\alpha}^4) \right] \right\}
\equiv \tilde{\eta}(\tilde{\alpha}),
\ee
where we have used Eq.(\ref{eq:exp}).

In the limit of weak lensing, both the modified reduced convergence
$\tilde{\eta}$ and the reduced convergence $\eta$
approach the local smoothness parameter $\tilde{\alpha}$:
$\eta \simeq \tilde{\eta} \simeq \tilde{\alpha}$.

\section{Dispersion in the local smoothness parameter
due to inhomogeneous matter distribution}

It is most convenient to illustrate the effect of gravitational
lensing by comparing the angular diameter distance in
a smooth univerve (which is equal to Dyer-Roeder distance
with smoothness parameter $\tilde{\alpha}=1$ \citep{Wang99})
with Dyer-Roeder distances with $\tilde{\alpha} \neq 1$.
Since this is often done by the community,
it is important that we qualify such illustration by
noting the following:\\
\noindent
(1) A realistic description of weak lensing of point sources
using the smoothness parameter $\tilde{\alpha}$ {\it requires}
generalizing $\tilde{\alpha}$ to be a {\it local} parameter,
the local smoothness parameter \citep{Wang99}.\\
\noindent
(2) The distribution of the possible values of the
local smoothness parameter $\tilde{\alpha}$ is
directly related to matter distribution in our universe.\\
\noindent
(3) The variance of $\tilde{\alpha}$ depends on the matter 
power spectrum. It provides a quantitative guidance in
the use of Dyer-Roeder distances to 
illustrate the effect of gravitational
lensing of point sources. For example, instead of arbitrary values
of $\tilde{\alpha}$, one should use $\tilde{\alpha}$
values which are one or two standard deviations
away from the mean value of $\langle \tilde{\alpha}\rangle=1$.

Next, we derive the variance of $\tilde{\alpha}$
in terms of the matter power spectrum.
This is done through the relation of
$\tilde{\alpha}$ to the reduced convergence
$\eta$ derived in Sec.III.
 
In Sec.III, we have shown that $\tilde{\kappa}_{min}$ is the
true minimum convergence and corresponds 
to the minimum magnification $\mu_{min}$, while
$\kappa_{min}$ is the approximation to $\tilde{\kappa}_{min}$ 
in the weak lesning limit.
This suggests that we need to use 
$\tilde{\eta}=1+\kappa/|\tilde{\kappa}_{min}|$, 
instead of $\eta=1+\kappa/|\kappa_{min}|$,
as the reduced convergence. Subsequently, we
modify the formulae in \cite{Valageas00}
as follows.

The convergence $\kappa$ \citep{Bernardeau97,Kaiser98}
is modified to be
\be
\label{eq:kappamod}
\kappa=\frac{3}{2}\, \Omega_m\, cH_0^{-1}\, \int_0^{z_s} 
dz\, \frac{\tilde{w}(z, z_s)}{E(z)}\, \delta,
\ee
where $z_s$ is the source redshift, $\delta=
(\rho-\overline{\rho})/\overline{\rho}$, and
\be
\tilde{w}(z, z_s)= \left(\frac{H_0}{c} \right)^2\,
\frac{r(z)}{r(z_s)}\, (1+z)^2 (1+z_s)\,
\left[ \lambda(z_s)-\lambda(z) \right].
\ee
The minimum of $\kappa$ occurs when $\rho=0$.
It is straightforward to check that the minimum
of the $\kappa$ as given in Eq.(\ref{eq:kappamod}) 
is $\tilde{\kappa}_{min}$ (see Eq.(\ref{eq:kapdef})),
which does correspond to
the minimum magnification (see Eq.(\ref{eq:mumin})). 

The variance of $\tilde{\eta}$ is similar to that of
$\eta$ \citep{Valageas00}, and given by
\be
\label{eq:xieta}
\xi_{\tilde{\eta}} = \left( \frac{3\Omega_m}{2} \right)^2\,
cH_0^{-1} \int_0^{z_s} \mathrm{d}z\, \frac{ \tilde{w}^2(z, z_s)}{E(z)}\,
\frac{I_{\mu}(z)}{|\tilde{\kappa}_{min}(z)|^2},
\ee
with $\tilde{\kappa}_{min}(z)$ given by Eq.(\ref{eq:kapdef}), and
\be
I_{\mu}(z)= \pi \int_0^{\infty} \frac{\mathrm{d}k}{k}\,\,
\frac{\Delta^2(k,z)}{k}\, W^2({\cal D}k\theta_0),
\ee
where $\Delta^2(k,z)= 4\pi k^3 P(k,z)$, $k$ is the wavenumber,
and $P(k,z)$ is the matter power spectrum. 
The window function $W({\cal D}k\theta_0)=2J_1({\cal D}k\theta_0)/
({\cal D}k\theta_0)$ for smoothing angle $\theta_0$. Here $J_1$ is the Bessel 
function of order 1.

Since $\tilde{\eta}$ and $\tilde{\alpha}$ are related 
(see Eq.(\ref{eq:etatil-alpha})), we have
\be
\xi_{\tilde{\eta}}  = \lim_{N \rightarrow \infty} \frac{1}{N} \sum_1^N
\left( \tilde{\eta}_i - \langle \tilde{\eta} \rangle \right)^2
=\left[ \frac{\partial \tilde{\eta} }{\partial \tilde{\alpha}} \right]^2 
\xi_{\tilde{\alpha}},
\ee
where $\xi_{\tilde{\alpha}}$ is the variance of $\tilde{\alpha}$.
Hence
\be
\label{eq:xialpha}
\xi_{\tilde{\alpha}} = \xi_{\tilde{\eta}} 
\left[ \frac{\partial \tilde{\eta} }{\partial \tilde{\alpha}} \right]^{-2},
\ee
with $\tilde{\eta}(\tilde{\alpha})$ given by Eq.(\ref{eq:etatil-alpha}),
and $\xi_{\tilde{\eta}}$ given by Eq.(\ref{eq:xieta}).

\section{Summary and discussions}

We have related the local smoothness
parameter $\tilde{\alpha}$ to the
reduced convergence $\eta= 1+ \kappa/|\kappa_{min}|$. 
This establishes the connection between two different
approached to the weak lensing of point sources.

For simplicity and convenience,
it is likely that the community (especially observers)
will continue to use Dyer-Roeder distances
for various values of the smoothness parameter
$\tilde{\alpha}$ to illustrate the effect of weak lensing 
on point sources. In order to provide meaningful guidance to such
illustrations, we have devived the variance 
of the local smoothness parameter
$\tilde{\alpha}$ in terms of the matter power spectrum
(see Eqs.(\ref{eq:xialpha}), (\ref{eq:xieta}),
and (\ref{eq:etatil-alpha})).

Wang, Holz, \& Munshi (2002) have used $\eta$
to derive a universal probability distribution (UPDF)
for weak lensing amplification. 
We propose a corrected definition
for the reduced convergence: $\tilde{\eta}= 1+ 
\kappa/|\tilde{\kappa}_{min}|$, with $\mu_{min}
=1/(1-\tilde{\kappa}_{min})^2$. 
We have shown that the true minimum convergence is $\tilde{\kappa}_{min}$
(defined as the convergence corresponding to the minimum magnification
$\mu_{min}$), while $\kappa_{min}(z)$ (as commonly defined in weak lensing
literature) is only its approximation when the filled beam (instead of the
empty beam) angular diameter distance is used along the line of sight
from the observer to the source.

Note that $\kappa_{min}$ does {\it not} correspond to $\mu_{min}$.
However, the correction needed for $\kappa_{min}$ is modest;
$\kappa_{min}$ is smaller than $\tilde{\kappa}_{min}$ by less than 2\%
in a $\Omega_m=0.3$ universe at $z<1$ (see Fig.2).
Thus we don't expect dramatic corrections to 
the various published weak lensing statistics.

As a by-product of our derivation relating 
the local smoothness parameter $\tilde{\alpha}$
and the reduced convergence $\tilde{\eta}$,
we obtained simple and accurate analytical expressions
for the angular diameter distance, $D_A(\tilde{\alpha}|z)$, 
which are solutions to the Dyer-Roeder equation.
We find an expression of $D_A(\tilde{\alpha}|z)$ as a polynomial
in the local smoothness parameter $\tilde{\alpha}$ (with 
coefficients given by integrals) that can be made
arbitrarily accurate, and is valid for {\it all} cosmological models
(see Eq.(\ref{eq:exp}) in Sec.{\ref{app:DAalpha}). 
In the Appendix, we give simple analytical approximations of 
the coefficients in Eq.(\ref{eq:exp}) for flat models with a cosmological
constant, and open models without dark energy.
These analytical approximations give angular diameter
distances that have an accuracy 
better than 0.1\% up to a redshift of 2.

Our results should be useful in
studying the weak lensing of standard candles.

\acknowledgements
We thank Ron Kantowski for helpful discussions,
This work was supported in part by 
NSF CAREER grant AST-0094335, and the NSF REU
program at the University of Oklahoma.

\appendix
\section{Analytical Approximations
for $\tilde{\kappa}_{min}(z)$ and $C_i(z)$}

Eq.(\ref{eq:exp}) gives the Dyer-Roeder distance 
in terms of a polynomial in the local smoothness parameter
$\tilde{\alpha}$. 
For clarity and convenience, let us now derive 
analytical approximations for $\tilde{\kappa}_{min}(z)$,  
and $C_i(z)$ ($i=1,2,3,4$), for
the special case of the dark energy being a cosmological constant.

First, let us expand the 
integral expressions in Eqs.(\ref{eq:kapdef}) and (\ref{eq:A,B,C,D})
in the limit of small $z$. We find for $z\ll 1$,
\ba
\tilde{\kappa}_{min}(z) & \simeq & - \frac{\Omega_m}{4}\, z^2\, 
\left[1 - 0.5\, (1+2 q_0)\,z\right], \nonumber \\
C_1(z) & \simeq & \frac{3}{40}\, \Omega_m\, z^2\, 
\left[1 - 0.5\, (1+2 q_0)\, z\right], \nonumber \\
C_i(z) & \simeq & \overline{C_i}\, \Omega_m^2\, z^{2i}\, 
\left[1 - 0.5\,i\, (1+2 q_0)\, z\right], \hskip 1cm (i=2,3,4)
\ea
where $\overline{C_2}=\frac{3}{1120}$,
$\overline{C_3}=\frac{1}{17920}$, 
$\overline{C_4}=\frac{3}{3942400}$, and
$q_0 \equiv \Omega_m/2 - \Omega_{\Lambda}$.

$C_i(z)$ ($i=1,2,3,4$) can be rewritten as
\ba
C_1(z) & \equiv &C_{10}(z) \,\left|\tilde{\kappa}_{min}(z) \right|, \nonumber\\
C_i(z) & \equiv &C_{i0}(z) \, C_1(z) C_{i-1}(z), \hskip 1cm (i=2,3,4)
\ea
where $C_{i0}$ ($i=1,2,3,4$) are nearly constant. We find that 
\be
\label{eq:B0,C0,D0}
C_{10}(z) \simeq 0.3, \hskip 1cm
C_{20}(z) \simeq \frac{10}{21}, \hskip 1cm
C_{30}(z) \simeq \frac{5}{18}, \hskip 1cm
C_{40}(z) \simeq \frac{2}{11}.
\ee

Guided by the small $z$ limit, we are able to find
simple analytical forms to approximate 
$\tilde{\kappa}_{min}(z)$ and $A_0(z)$. 

For a flat universe, we found:
\ba
\label{eq:flatkap}
\tilde{\kappa}_{min}(z) &=& -\frac{\Omega_m}{4}\, \frac{z^2}{E(z)}\,
\left[ a_1 \, \ln\left( a_2 z^{a_3} +1 \right) +1 \right], \nonumber\\
a_1(\Omega_m)&= &.06354\,\Omega_m^{-1.5498}+.2688\Omega_m^{5.8586}+1.5812,
\nonumber \\
a_2(\Omega_m) &=& 3.6049\,\left( \Omega_m^{.7189}e^{.1082 
\Omega_m}-\Omega_m \right)-.1751,
\nonumber \\
a_3(\Omega_m) &=& \Omega_m^{.1306} e^{.6022\Omega_m}-1.4427\,\Omega_m+.5472.	  
\ea
and
\be
\label{eq:flatA0}
C_{10}(z)=0.3+\left(-0.00548\, \Omega_m^{1.57}-0.02725\right)
\, \ln\left[ E(z)- z\, \left( 1+ q_0 \right) \right].
\ee

For an open universe, we find
\be
\label{eq:openkap}
\tilde{\kappa}_{min}(z)= - \frac{\Omega_m}{4} \,
\frac{z^2\,(1+0.5\,z)}{E(z)}\,
\left\{ 1 -\frac{0.19}{\Omega_m^{1.08}}\,
\frac{\ln\left[E(z)- z\, \left( 1+ q_0\right) \right]}
{ z^{0.45\, \left(|\ln\Omega_m| +0.04 \right)^{0.7}}}
\right\},
\ee		
and 
\be 
\label{eq:openA0}		
C_{10}(z)=0.3+\left[\Omega_m^{0.252}\,\ln\left(\Omega_m^{0.0557}-
0.284\right)+0.302\right]
\, \ln\left[ E(z)- z\, \left( 1+ q_0 \right) \right].
\ee

Figs.3-4 show the accuracy of our analytical approximations
for $\tilde{\kappa}_{min}(z)$ (see Eqs.(\ref{eq:flatkap})(\ref{eq:openkap}))
and $C_{10}(z)$ (see Eqs.(\ref{eq:flatA0})(\ref{eq:openA0}))
as function of $\Omega_m$, for redshifts up to 2 and 5 respectively.
Each figure displays the maximum of the ratio of the difference between our
analytical approximation and the exact expression over
the latter.

Fig.5(a) shows the difference between the angular diameter distance given
by the numerical solution of the Dyer-Roeder equation and our analytical expansion
to 4th order in $\tilde{\alpha}$ (see Eq.(\ref{eq:exp})), with
$\tilde{\kappa}_{min}(z)$ and $C_i(z)$ ($i=1,2,3,4$) given by 
Eqs.(\ref{eq:B0,C0,D0})-(\ref{eq:openA0}).
The accuracy shown in Fig.5(a) is primarily limited by
our approximations for $\tilde{\kappa}_{min}(z)$, Eqs.(\ref{eq:flatkap})
and (\ref{eq:openkap}).

Fig.5(b) is the same as Fig.5(a), except with $\tilde{\kappa}_{min}(z)$
given by Eq.(\ref{eq:kapdef}), with the empty and full beam
distances, $D_A(\tilde{\alpha}=0|z)$ and $D_A(\tilde{\alpha}=1|z)$,
computed by integrating the Dyer-Roeder equation.

Comparison of Figs.3-4 with Fig.5 shows that
the errors in estimated distances using
analytical approximations of Eqs.(\ref{eq:B0,C0,D0})-(\ref{eq:openA0})
are much smaller than the errors in $\tilde{\kappa}_{min}(z)$ and $A_0(z)$.
This is because
\be
\frac{\Delta D_A(\tilde{\alpha}|z)}{D_A(\tilde{\alpha}|z)}
\simeq \frac{\Delta |\tilde{\kappa}_{min}(z)|}{|\tilde{\kappa}_{min}(z)|}
\,\left| \frac{D_A(\tilde{\alpha}=1|z)}{D_A(\tilde{\alpha}|z)}
-1 \right|
\ll \frac{\Delta |\tilde{\kappa}_{min}(z)|}{|\tilde{\kappa}_{min}(z)|}.
\ee
The last inequality arises from the fact that
$D_A(\tilde{\alpha}=1|z)$ and $D_A(\tilde{\alpha}|z)$ differ
by less than $\sim\,$20\% for $\tilde{\alpha}\la 2$.

For all practical purposes in the weak lensing of standard candles,
the local smoothness parameter $\tilde{\alpha}\la 2$,
since the probability for $\tilde{\alpha}\ga 2$ 
is vanishingly small for all cosmological models
at all redshifts \citep{Wang99}.
Note that the probability distribution of $\tilde{\alpha}$
becomes {\it narrower} and peaked closer to $\tilde{\alpha}=1$ as redshift
$z$ increases, as the universe is smoother on the average at high $z$
\citep{Wang99}.
Our analytical approximations, given by
Eqs.(\ref{eq:exp}), (\ref{eq:B0,C0,D0})-(\ref{eq:openA0}), 
give an accuracy of better than about 0.1\% in all cosmological
for $\tilde{\alpha}\la 2$ (see Fig.5(a)).
If the exact expression for $\tilde{\kappa}_{min}(z)$
(see Eq.(\ref{eq:kapdef})),
instead of Eqs.(\ref{eq:flatkap}) and (\ref{eq:openkap}),
is used, the accuracy in the estimated distance is better than 0.02\% 
for all cosmological models 
for $\tilde{\alpha}\la 2$ (see Fig.5(b)).

\begin{figure}
\plotone{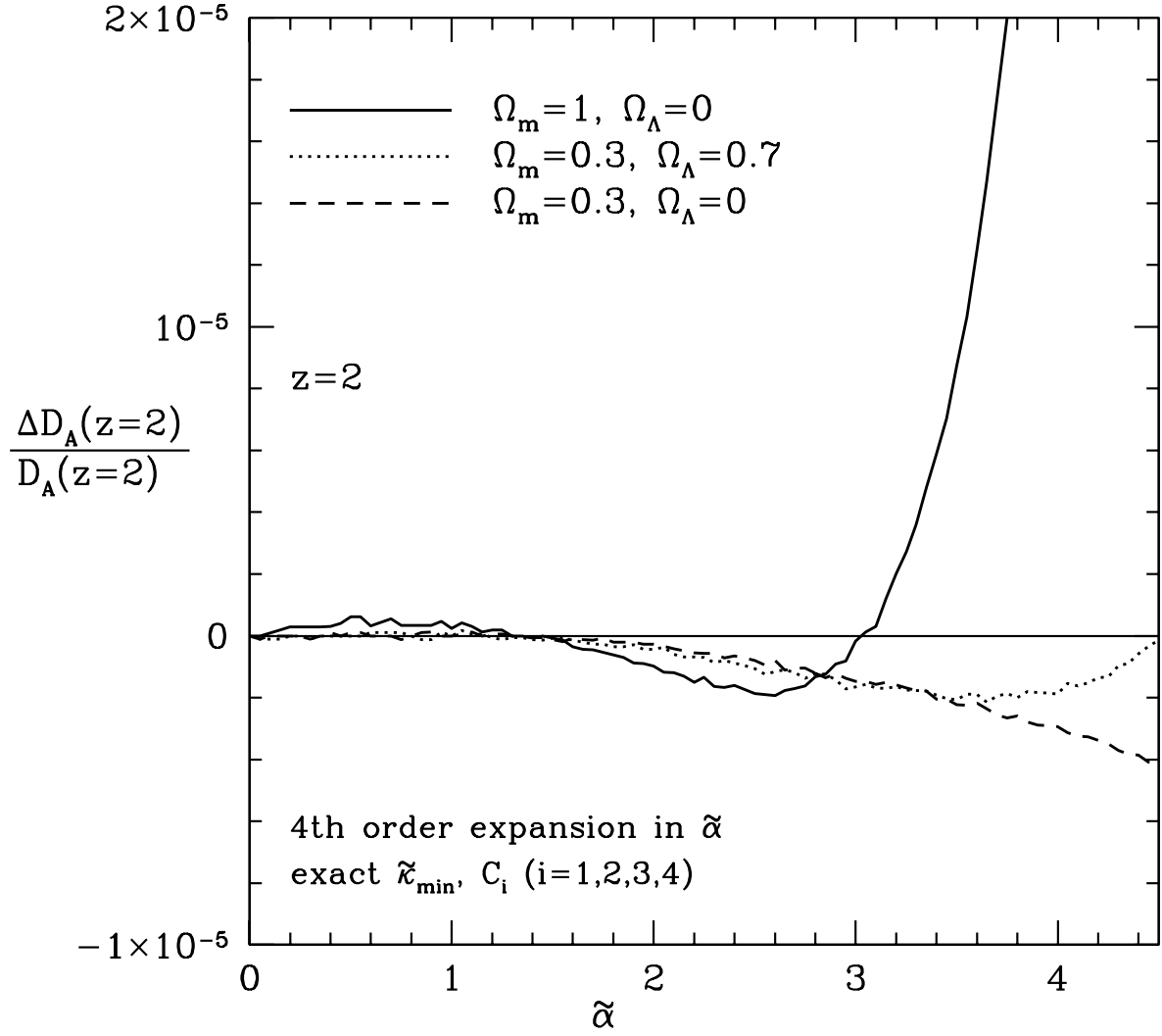}
\figcaption{
\label{fig:DAalpha} The difference between the angular diameter distance given
by the solution of the Dyer-Roeder equation and our analytical expansion
to 4th order in $\tilde{\alpha}$ (see Eq.(\ref{eq:exp})).
}
\end{figure}

\begin{figure}
\plotone{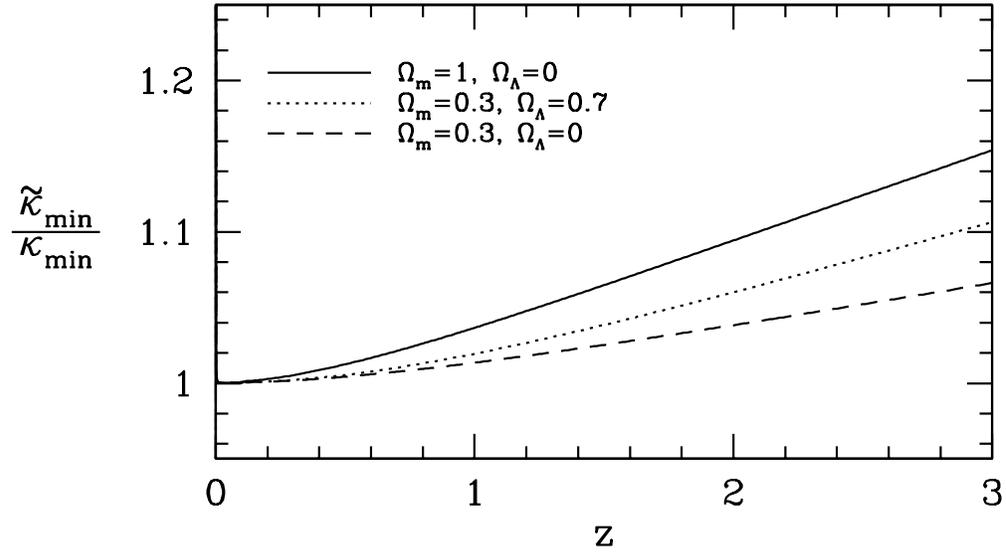}
\figcaption{The difference between $\tilde{\kappa}_{min}(z)$
and $\kappa_{min}(z)$ for three representative cosmological models.
}
\end{figure}

\begin{figure}
\plotone{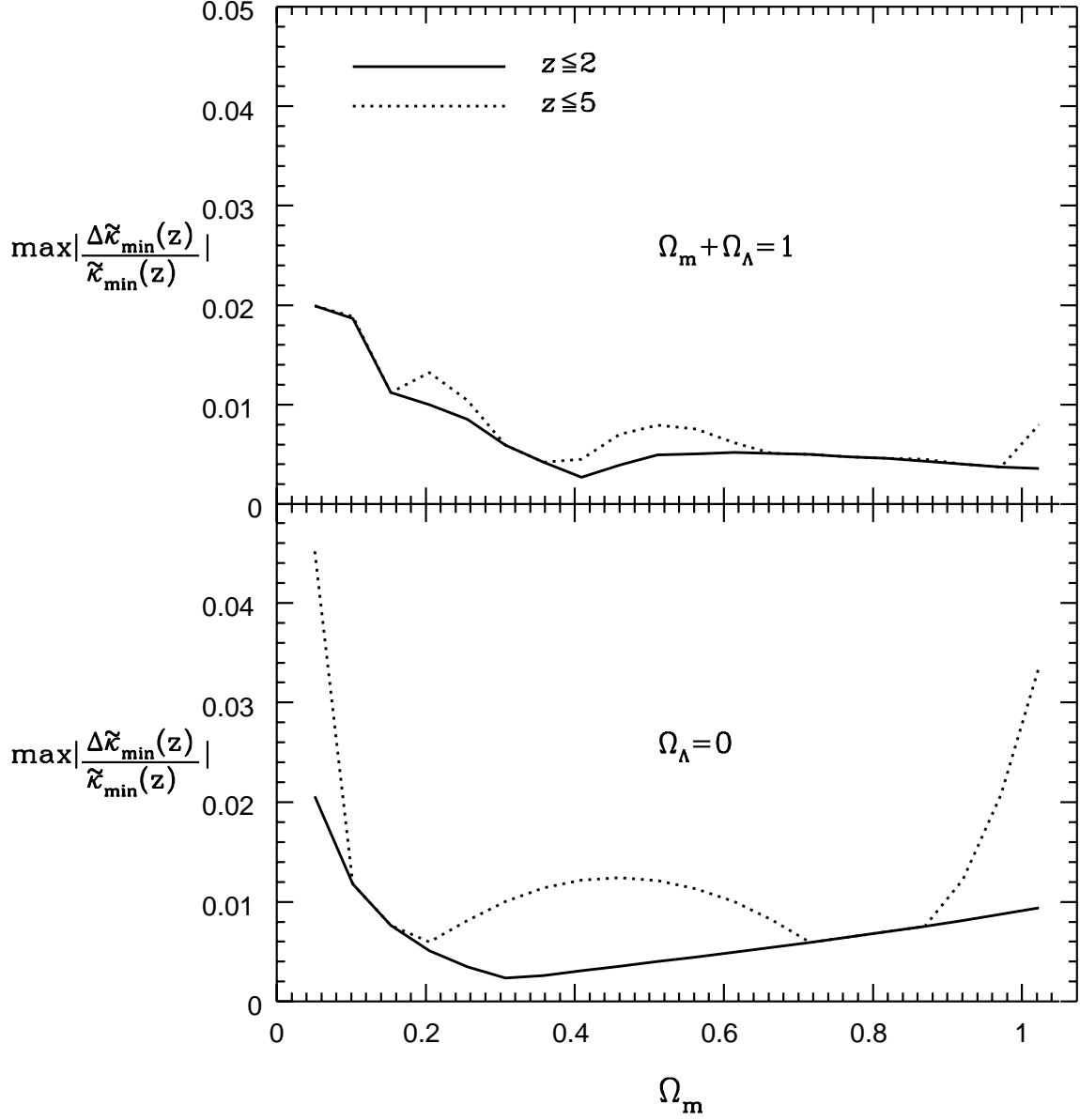}
\figcaption{The accuracy of our analytical approximations
for $\tilde{\kappa}_{min}(z)$ (see Eqs.(\ref{eq:flatkap})(\ref{eq:openkap})).
}
\end{figure}

\begin{figure}
\plotone{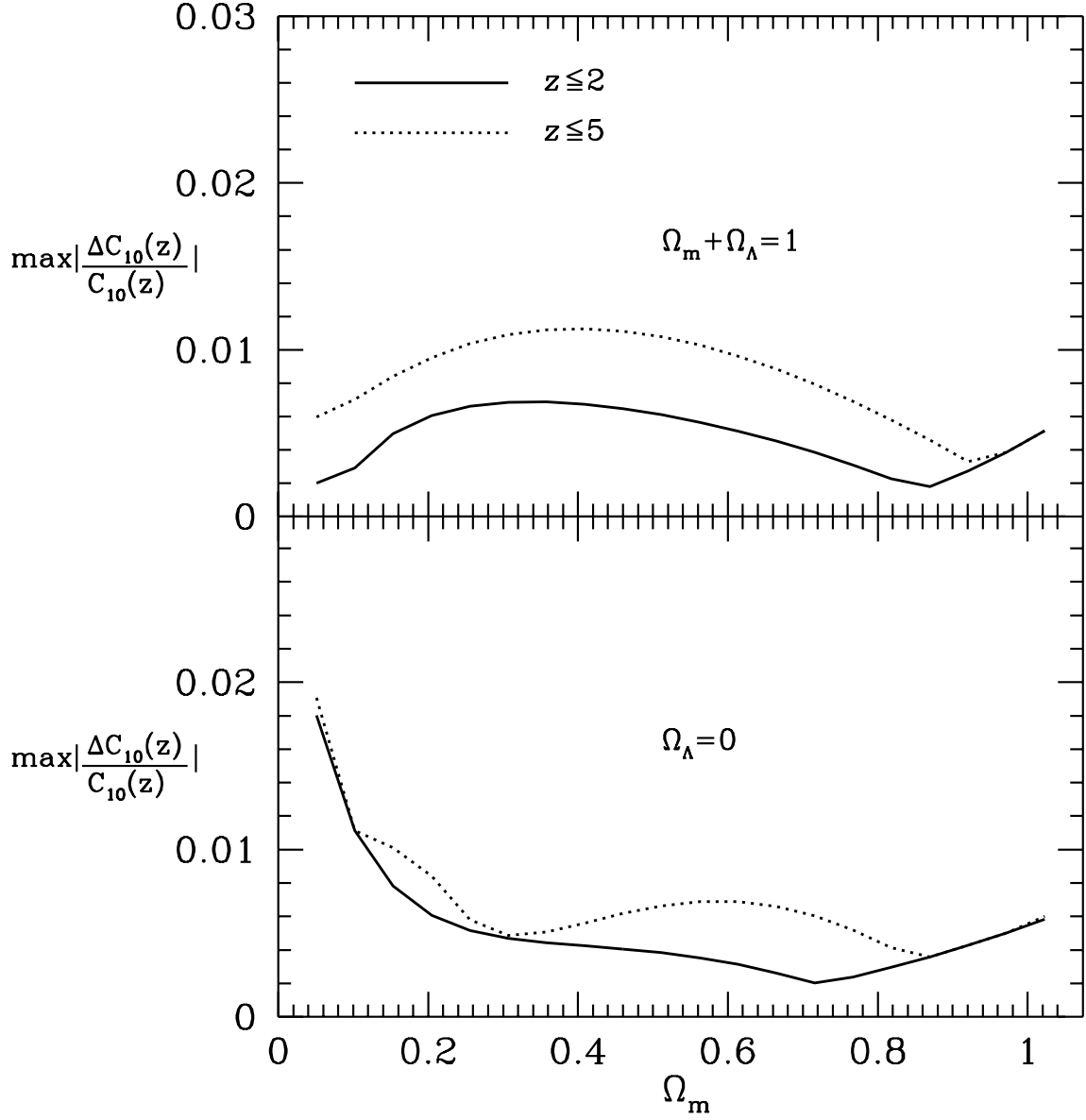}
\figcaption{The accuracy of our analytical approximations
for $C_{10}(z)$ (see Eqs.(\ref{eq:flatA0})(\ref{eq:openA0})).}
\end{figure}

\begin{figure}
\plotone{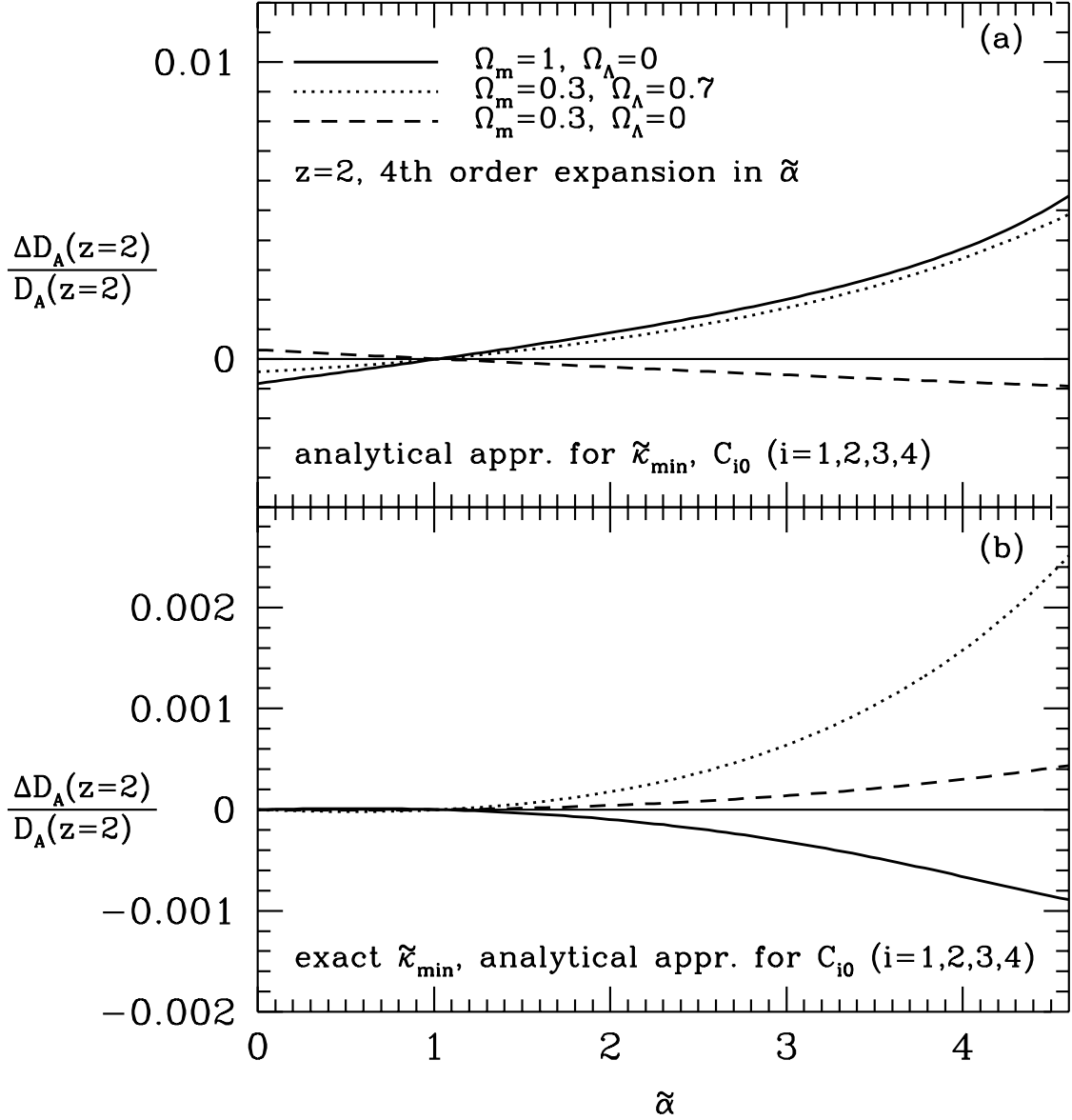}
\figcaption{(a) The difference between the angular diameter distance given
by the solution of the Dyer-Roeder equation and our analytical expansion
to 4th order in $\tilde{\alpha}$ (see Eq.(\ref{eq:exp})), with
$\tilde{\kappa}_{min}(z)$, and $C_i(z)$ ($i=1,2,3,4$) given by 
Eqs.(\ref{eq:B0,C0,D0})-(\ref{eq:openA0}).
(b) The same as (a), except with $\tilde{\kappa}_{min}(z)$
given by Eq.(\ref{eq:kapdef}), with the empty and full beam
distances, $D_A(\tilde{\alpha}=0,z)$ and $D_A(\tilde{\alpha}=1,z)$,
computed by integrating the Dyer-Roeder equation.}
\end{figure}

\end{document}